\documentclass{nature}

\usepackage{graphicx} 
\usepackage[english]{babel}
 \usepackage{url}
 \usepackage{graphics}

\title{Assembling large, complex environmental metagenomes}

\author{Adina Chuang Howe$^{1,2}$, 
Janet Jansson$^{3,4}$,
Stephanie A. Malfatti$^{3}$,
Susannah G. Tringe$^{3}$,
James M. Tiedje$^{1,2}$, and 
C. Titus Brown$^{1,5\ast}$}

\begin{document}

\maketitle

\begin{affiliations}
\item Microbiology and Molecular Genetics, Michigan State University, East Lansing, MI, USA\\
\item Plant, Soil, and Microbial Sciences, Michigan State University, East Lansing, MI, USA\\
\item Department of Energy (DOE) Joint Genome Institute, Walnut Creek, CA, USA\\
\item Lawrence Berkeley National Laboratory, Earth Sciences Division, Berkeley, CA, USA\\
\item Computer Science and Engineering, Michigan State University, East Lansing, MI, USA\\
\end{affiliations}

\begin{abstract}
The large volumes of sequencing data required to sample complex
environments deeply pose new challenges to sequence analysis approaches.
\emph{De novo} metagenomic assembly effectively reduces the total
amount of data to be analyzed but requires significant computational
resources.  We apply two pre-assembly filtering approaches, digital
normalization and partitioning, to make large metagenome assemblies more computationaly tractable.  Using a human gut
mock community dataset, we demonstrate that these methods result in
assemblies nearly identical to assemblies from unprocessed data.  We
then assemble two large soil metagenomes from matched Iowa corn and
native prairie soils.  The predicted functional content and phylogenetic
origin of the assembled contigs indicate significant taxonomic
differences despite similar function.  The assembly strategies
presented are generic and can be extended to any metagenome; full
source code is freely available under a BSD
license.
\end{abstract}

Complex microbial communities operate at the heart of many crucial
terrestrial, aquatic, and host-associated processes, providing
critical ecosystem functionality that underpins much of biology
\cite{Arumugam:2011p735,Hess:2011p686,Iverson:2012p1281,
  Mackelprang:2011p1087,Qin:2010p189,Tringe:2005p174,Venter:2004p170}.
These systems are difficult to study {\em in situ}, and we are only
beginning to understand their diversity and functional potential.
Advances in DNA sequencing now provide unprecedented
access to the genomic content of these communities via shotgun
sequencing, which produces millions to billions of short-read
sequences \cite{Hess:2011p686,Mackelprang:2011p1087,Qin:2010p189}.
Because shotgun sequencing samples communities randomly, ultradeep
sequencing is needed to detect rare species in environmental samples,
with an estimated 50 Tbp needed for an individual gram of soil
\cite{Gans:2005p1365}.  Both short read lengths and the large volume
of sequencing data pose new challenges to sequence analysis
approaches.  A single metagenomic project can readily generate as much
or more data than is in global reference databases; for example, a
human-gut metagenome sample containing 578 Gbp \cite{Qin:2010p189}
produced more than double the data in NCBI RefSeq
(Release 56).  Moreover, short reads contain only minimal
signal for homology searches and are error-prone, limiting direct
annotation approaches against reference databases.  And finally, the
majority of genes sequenced from complex metagenomes typically contain little or no
similarity to experimentally studied genes, further complicating
homology analysis \cite{Arumugam:2011p735,Qin:2010p189}.

\emph{De novo} assembly of raw sequence data offers several advantages
over analyzing the sequences directly.  Assembly removes most random
sequencing errors and decreases the total amount of data to be
analyzed.  These resulting assembled contigs are longer than
sequencing reads and provide gene order.  Importantly, \emph{de novo}
assembly does not rely on the existence of reference genomes, thus
allowing for the discovery of novel genomic elements.  The main challenge for
metagenomic applications of \emph{de novo} assembly is that current
assembly tools do not scale to the high diversity and large volume of metagenomic data: metagenomes from rumen, human gut, and permafrost
soil sequencing could only be assembled by discarding low abundance
sequences prior to assembly
\cite{Hess:2011p686,Mackelprang:2011p1087,Qin:2010p189}.  Traditional
assemblers are designed for single genomes whose abundance
distribution and diversity content are typically simpler than
community metagenomes.  Although many metagenome-specific assemblers
have recently been developed for community assembly, most of these assemblers
do not scale to extremely large samples \cite{Scholz:2012p1372}.

Here, we combine two approaches, digital normalization and
partitioning, to tackle the problem of {\em de novo} metagenome
assembly.  Digital normalization normalizes sequence coverage and
reduces the dataset size by discarding reads from high-coverage
regions \cite{browndiginorm}.  Subsequently, partitioning separates
reads based on transitive connectivity, resulting in easily assembled
subsets of reads \cite{howeartifacts,Pell:2012cq}.  We evaluate these
approaches by applying them to a human gut mock community dataset, and
find that these filtering methods result in assemblies nearly
identical to assemblies from the unprocessed dataset.  Next, we apply
these approaches to the assembly of metagenomes from two matched soils,
100-year cultivated Iowa agricultural soil and native Iowa
prairie.  We compare the predicted functional capacities and
phylogenetic origins of the assembled contigs and conclude that
despite significant phylogenetic differences, the functions encoded in
both soil data sets are similar.  We also show that virtually no
strain-level heterogeneity is detectable within the assembled reads.

\section*{Results}

\section*{Data reduction results in similar assemblies}

We evaluated the recovery of reference genomes from {\em de novo}
metagenomic assembly by comparing unfiltered traditional assembly to
the the described filtered assembly (Fig. \ref{flowchart}; see Methods and
Supplementary Information). Initially, the abundance of genomes within the mock
dataset was estimated based on the reference genome coverage of
sequence reads in the unfiltered dataset.  Coverage (excluding genomes
with less than 3-fold coverage) ranged from 6-fold to 2,000-fold
(Supplementary Table 1 and Supplementary Fig. 2 and 3).  Overall, the
unfiltered dataset reads covered a total of 93\% of the reference
genomes.  Filtering removed 5.9 million reads, 40\% of the total (Table ~\ref{data-summary});  the remaining
reads covered 91\% of the reference genomes
(Table ~\ref{data-summary} and Supplementary Fig. 2 and 3).

We next compared the recovery of reference genomes in contigs
assembled from the original and filtered datasets.  Using the Velvet
assembler, we recovered 43\% and 44\% of the reference genomes, respectively.  The
assembly of the original dataset contained 29,063 contigs and 38
million bp, while the filtered assembly contained 30,082 contigs and
35 million bp (Table ~\ref{assembly-summary}).  Comparable recoveries
of references between original and filtered datasets were also
obtained with other assemblers (SOAPdenovo and Meta-IDBA, 
Table ~\ref{assembly-summary}).  Overall,
the unfiltered and filtered assemblies were very similar, sharing 95\%
of genomic content.  For the highest abundance references
(the plasmids NC\_005008.1, NC\_005007.1, and
NC\_005003.1), the unfiltered assembly recovered
significantly more of the original sequence; however, for the large
majority of genomes, the filtered assembly recovered similar (and
sometimes greater) amounts of the reference genomes (Supplementary Fig. 2 and
3).  The distribution of contig lengths in unfiltered and filtered
assemblies were also comparable (Supplementary Fig. 4).

We estimated the abundance of assembled contigs and reference genomes
using the mapped sequencing reads (Supplementary Fig. 5).  Above a
sequencing coverage of five, the majority of reads which could be
mapped to reference genomes were included in the assembled contigs
(Supplementary Fig. 3 and 4).  Below this threshold, reads could be mapped to
reference genomes but were less likely to be associated with assembled
contigs.  We next compared the abundances of the reference genomes to
the abundances of the contigs in the unfiltered and filtered
assemblies.  The abundance estimations from the filtered assembly were
significantly closer to predicted abundances from reference genomes
(\emph{n} = 28,652; p-value = 0.032, see Supplementary Information).

\section*{Partitioning separates most reads by species}

We next partitioned the filtered data set based on de Bruijn graph
connectivity and assembled each partition independently
\cite{howeartifacts, Pell:2012cq}. The resulting assemblies of
unpartitioned and partitioned were more than 99\% identical.  In the
mock dataset, we identified 9 million reads in 85,818 disconnected
partitions (Supplementary Fig. 6).  Among these, only 2,359 (2.7\%) of the
partitions contained reads originating from more than one genome,
indicating that partitioning separated reads from distinct
species.

In general, reference genomes with high sequencing coverage were
associated with fewer partitions (Supplementary Table 1): a
total of 112 partitions contained reads from high abundance reference
genomes (coverage above 25) compared to 2,771 partitions associated
with lower abundance genomes.  This is consistent
with the observation that the main effect of low coverage is to
``break'' connectivity in the assembly graph \cite{Chaisson:2008p1373,Pevzner:2001p1374}.

To further evaluate the effects of partitioning, we introduced spiked reads from
\emph{E. coli} genomes into the mock community
dataset. First, simulated reads from a single genome (\emph{E. coli} strain
E24377A, NC\_009801.1 with 2\% substitution error and 10x coverage) were added to the
mock community dataset and then processed in the same way as the
unfiltered mock dataset.  We observed similar amounts of data
reduction after digital normalization and partitioning (Table
~\ref{data-summary}).  Among the 81,154 partitioned sets of reads, we
identified only 2,580 (3.2\%) partitions containing reads from
multiple genomes.  A total of 424 partitions contained reads from the
spiked \emph{E. coli} genome (201 partitions contained \emph{only}
spiked reads) and when assembled aligned to 99.5\% of \emph{E. coli}
strain E24377A genome (4,957,067 of 4,979,619 bp) (Supplementary Fig. 6).

Next, we introduced five closely-related \emph{E. coli} strains into the mock
community dataset and performed the same analysis.
Partitioning this ``mix-spiked" mock community
dataset resulted in 81,425 partitions, of which 1,154 (1.4\%)
partitions contained reads associated with multiple genomes.  Among
the partitions which contained reads associated with a single genome,
658 partitions contained reads originating from one of the spiked
\emph{E. coli} strains.  In partitions containing reads from more than
one genome, 224 partitions contained reads from a spiked
\emph{E. coli} strain and one other reference genome (either another
spiked strain or from the mock community dataset) (Supplementary Fig. 7).  We
independently assembled the partitions containing reads originating
from the spiked \emph{E. coli} strains.  Among the resulting 6,076
contigs, all but three contigs originated from a spiked \emph{E. coli}
genome.  The remaining three contigs were more than 99\% similar to
HMP mock reference genomes (NC\_000915.1, NC\_003112.2, and
NC\_009614.1).  The contigs associated with \emph{E. coli} were
aligned against the spiked reference genomes, recovering greater than
98\% of each of the five genomes.  Many of these contigs contained
similarities to reads originating from multiple genomes (Supp Fig. 8),
and 3,075 contigs (51\%) could be aligned to reads which originated from
more than one spiked genome.  This result is comparable to the
fraction of contigs which are associated with multiple genomes in the
unfiltered data set, where 66\% of 4,702 contigs associated with
spiked reads contain reads that originate from more than one spiked
genome.

\section*{Data reduction and partitioning enable the assembly of two soil metagenomes}

We next applied these approaches to the {\em de novo} assembly of two
soil metagenomes.  Iowa corn and prairie datasets (containing 1.8
billion and 3.3 billion reads, respectively) could not be assembled by
Velvet in 500 GB of RAM.  A 75 million reads subset of the Iowa corn
dataset alone required 110 GB of memory, suggesting that assembly of
the 3.3 billion read data set might need as much as 4 TB of RAM
(Supplementary Fig. 9).  Applying the same filtering approaches as described
above, the Iowa corn and prairie datasets were reduced to 1.4 billion
and 2.2 billion reads, respectively, and after partitioning, a total
of 1.0 billion and 1.7 billion reads remained, respectively.  Prefiltering
used 300 GB of RAM or less.  The
large majority of k-mers in the soil metagenomes are relatively
low-abundance (Fig. ~\ref{diginormcoverage}), and consequently digital
normalization did not remove as many reads in the soil metagenomes as
in the mock data set.

Based on the mock community dataset, we estimated that above a
sequencing depth of five, the large majority of sequences could be
assembled into contigs larger than 300 bp (Supplementary Fig. 1).  Given the
greater diversity expected in the soil metagenomes, we normalized
these datasets to a sequencing depth of 20 (i.e., discarding redundant 
reads within dataset above this coverage).  
After partitioning the filtered datasets, we identified a total
31,537,798 and 55,993,006 partitions (containing more than five reads)
in the corn and prairie datasets, respectively.  For assembly, we
grouped partitions together into files containing 10 million reads.
Data reduction and partitioning were completed in less than 300 GB of
RAM; once partitioned, each group of reads could be assembled in less
than 14 GB and 4 hours.  This readily enabled the usage of multiple
assemblers and assembly parameters.

The final assembly of the corn and prairie soil metagenomes resulted
in a total of 1.9 million and 3.1 million contigs greater than 300 bp,
respectively, and a total assembly length of 912 million bp and 1.5
billion bp, respectively.  To estimate abundance of assembled contigs
and evaluate incorporation of reads, all quality-trimmed reads were
aligned to assembled contigs.  Overall, for the Iowa corn assembly,
8\% of single reads and 10\% of paired end reads mapped to the
assembly.  Among the paired end reads, 95.5\% of the reads aligned
concordantly.  In the Iowa prairie assembly, 10\% of the single reads
and 11\% of the paired end reads aligned to the assembled contigs, and
95.4\% of the paired ends aligned concordantly (Table
~\ref{read-map}).  Based on these mappings, we calculated read
recovery in assembled contigs within the soil metagenomes
(Fig. ~\ref{soilassemblycoverage}).  Overall, there is a positively
skewed distribution of read overage of all contigs from both soil
metagenomes, biased towards a coverage of less than ten-fold, and 48\%
and 31\% of total contigs in Iowa corn and prairie assemblies
respectively had a median basepair coverage less than 10.

Among contigs, the presence of polymorphisms was examined by
identifying the amount of consensus obtained by reads mapped
(Supplementary Information Methods).  For both the Iowa corn and prairie metagenomes,
more than 99.9\% of contigs contained base calls which were supported
by a 95\% consensus from mapped reads over 90\% of their lengths,
demonstrating an unexpectedly low polymorphism rate
(Supplementary Fig. 10).

\section*{Annotation of the soil assemblies revealed similar functional
profiles but different taxonomy}
 
We annotated assembled contigs through the MG-RAST pipeline, which
was modified to account for per-contig abundance.
This annotation resulted in 2,089,779 and 3,460,496 predicted protein
coding regions in the corn and prairie metagenomes, respectively.  The
large majority of these regions did not share similarity with any gene
in the M5NR database -- 61.8\% in corn and
70.0\% in prairie.  In total, 613,213 (29.3\%) and 777,454 (22.5\%)
protein coding regions were assigned to functional categories.  The
functional profile of these annotated features against SEED subsystems
were compared (Fig. ~\ref{subsystem}).  For both the corn and prairie
metagenomes, the most abundant functions in the assembly were
associated with the carbohydrate (e.g., central carbohydrate
metabolism and sugar utilization), amino acid (e.g., biosynthesis and
degradation), and protein (e.g., biosynthesis, processing, and
modification) metabolism subsystems.  The subsystem profile of both
metagenomes were very similar while the taxonomic profile of the
metagenomes based on the originating taxonomy (phyla) was different
(Fig. ~\ref{phyla}, Supp Methods).  Within both metagenomes,
Proteobacteria were most abundant.  In Iowa
corn, Actinobacteria, Bacteroidetes, and Firmicutes were
the next most abundant, while in the Iowa prairie, Acidobacteria,
Bacteroidetes, and Actinobacteria were the next most abundant.
The Iowa prairie also had nearly double the fraction
of Verrucomicrobia than did Iowa corn.

\section*{Discussion}


The diversity and sequencing depth represented by the mock community
is extremely low compared to that of most environmental metagenomes;
however, it represents a simplified, unevenly sampled model for a
metagenomic dataset which enables the evaluation of analyses through
the availability of source genomes.  For this dataset, the filtering
approaches described above were effective at reducing the dataset size
without significant loss of assembly.  This strategic filtering takes
advantage of the observed coverage ``sweet spot'' at which point
sufficient sequences are present for robust assembly (Supp Fig. 1).  
The normalization of sequences also resulted in
more even coverage (Fig. ~\ref{diginormcoverage}),
minimizing assembly problems caused by variable coverage.
Additional reduction of the dataset was achieved by the removal of
high abundance sequences \cite{howeartifacts}.

The specific effects of filtering varied depending on differences
between reference genomes.  Variable abundance and conserved regions
in references had an impact on filtered assembly recovery.  The
filtered assemblies of the three plasmids of the \emph{Staphylococcus
  epidermidis} genome (NC\_005008.1, NC\_005007.1, and NC\_005003.1)
were highly abundant (Supplementary Table 1) and shared several conserved
regions (90\% identity over more than 290 bp).  During normalization,
repetitive elements in these genomes would appear as high coverage
elements and be removed, as evidenced by a large difference in the
number of reads associated with NC\_005008.1 in the unfiltered and
filtered datasets (supplementary Fig. 2). Consequently, the unfiltered dataset
contained more reads spanning these repetitive regions.  This most
likely enabled assemblers to extend the assembly of these sequences
and resulted in the observed increased recovery of these genomes in
the unfiltered assemblies. This result, though rare among the mock
reference genomes, identifies a shortcoming of our approach, and
indeed for most short-read assembly approaches, related to repetitive
regions and/or polymorphisms.  For the soil metagenomes our data
reduction may have caused some information loss in exchange for the
ability to assemble previously intractable data sets.  Evaluation of
the mock community dataset suggests that this information loss is
minimal overall and that our approaches result in a comparable
assembly whose abundance estimations are slightly improved.


Metagenomes contain many distinct genomes, which are largely
disconnected from each other but which sometimes share sequences due
to conservation or lateral transfer.  Our prefiltering approach
removes both common multi-genome elements as well as artificial
connectivity stemming from the sequencing process.
As shown above on the mock data set, the removal of these sequences
does not significantly alter the recovery of reference genomes through
{\em de novo} assembly: the resulting assemblies of unpartitioned and
partitioned datasets were nearly identical for the mock data.  The
large majority of these partitions contained reads from a single
reference genome, supporting our previous hypothesis that most
connected subgraphs contain reads from distinct genomes
\cite{Pell:2012cq}.  As expected, high abundance, well-sampled genomes
were found to contain fewer partitions and low abundance, under-sampled
genomes contained more partitions, due to fragmentation of the
assembly graph.

We further examined the recovery of sequences through partitioning by
computationally spiking in one or more \emph{E. coli} strains before
applying filtering and partitioning.  When we spiked in a single
\emph{E. coli} strain, we could reassemble 99\% of the original genome
(Supplementary Fig. 6).  When we spiked in five closely related strains,
we could recover the large majority of the genomic
content of these strains, albeit largely in chimeric contigs (Supplementary
Fig. 8).  This result is not unexpected, as
assemblies of the unfiltered dataset resulted in a slightly higher
fraction of assembled contigs associated with multiple references.
Overall, closely related sequences which result from either repetitive
or inter-strain polymorphisms challenge assemblers, and our
approaches are not specifically designed to target such regions.
However, the partitions resulting from our approach could provide a 
much-reduced subset of sequences to be
targeted for more sensitive assembly approaches for highly variable regions
(i.e. overlap-layout-consensus approaches or abundance binning
approaches \cite{Sharon:2012kx}).

One valuable result of partitioning is that it subdivides our datasets
into sets of reads which can be assembled with minimal
computational resources.  For the mock community dataset, this gain
was small, reducing unfiltered assembly at 12 GB and 4 hours to less
than 2 GB and 1 hour.  However, for the soil metagenomes, previously
impossible assemblies could be completed in less than a day and in
under 14 GB of memory enabling the usage of multiple assembly
parameters (e.g., k-length) and multiple assemblers (Velvet,
SOAPdenovo, and MetaIDBA).


The final assemblies of the corn and prairie soil metagenomes resulted
in a total of 1.9 million and 3.1 million contigs, respectively, and a
total assembly length of 912 million bp and 1.5 billion bp,
respectively -- equivalent to $\approx$ 500 \emph{E. coli} genomes worth of DNA.  We evaluated these assemblies based on paired-end concordance, which showed that the majority of the assembled contigs agreed with the raw sequencing data.  Overall,
there is a positively skewed distribution of abundance of all contigs
from both soil metagenomes, biased towards an abundance of less than
ten, indicative of the low sequencing coverage of these
metagenomes.

This study represents the largest published soil metagenomic sequencing effort
to date, and these assembly results demonstrate the enormous amount of
diversity within the soil.  Even with this level of sequencing,
millions of putative genes were defined for each metagenome, with
hundreds of thousands of functions.
More than half of the assembled contigs are not similar to anything in
known databases, suggesting that soil holds considerable unexplored
taxonomic and functional novelty.
Among the
protein coding sequences which were annotated, comparisons of the two
soil datasets suggests that the functional profiles are more similar
to one another than the complementing phylogenetic profiles.  This
result supports previous hypotheses that despite large diversity with
two different soil systems, the microbial community provides similar
overall function \cite{Girvan:2005jv,McGradySteed:1997uj,Muller:2002cd,Konstantinidis:2004hr}.



We present two strategies that readily enable the assembly of very
large environmental metagenomes by discarding redundancy and
subdividing the data prior to assembly.  These strategies are generic and should
be applicable to any metagenome.  We demonstrate their effectiveness by
first evaluating them on the assembly of a mock community metagenome,
and then applying them to two previously intractable soil metagenomes.

Partitioning is an especially valuable approach because it enables the
extraction of read subsets that should assemble together.  These read
partitions are small enough that a variety of assembly, abundance
analysis, and polymorphism analysis techniques can be easily applied
to them individually.

The two soil assemblies also provide a deeper glimpse of the
opportunities and challenges of large scale environmental metagenomics
in high-diversity systems such as soil: we identified millions of
novel putative proteins, most of unknown function.

\begin{addendum}
\item This project was supported by Agriculture and Food Research Initiative
Competitive Grant no. 2010-65205-20361 from the United States
Department of Agriculture, National Institute of Food and Agriculture
and National Science Foundation IOS-0923812, both to C.T.B.  A.H. was
supported by NSF Postdoctoral Fellowship Award \#0905961 and the Great
Lakes Bioenergy Research Center (Department of Energy BER
DE-FC02-07ER64494).  The work conducted by the U.S. Department of
Energy Joint Genome Institute is supported by the Office of Science of
the U.S. Department of Energy under Contract No. DE-AC02-05CH11231.
We acknowledge the support of Krystle Chavarria and 
Regina Lamendella for extraction of DNA from Great Prairie soil samples
and the technical support of Eddy Rubin and Tijana Glavina del Rio 
at the DOE JGI and John Johnson and Eric McDonald at MSU HPC.  
\item[Author Contributions] A.H. and C.T.B. designed experiments and wrote paper.
A.H. performed experiments and analyzed the data.  J.J., S.T., and J.T. discussed results 
and commented on manuscript.  S.M. managed raw sequencing datasets.
\item[Competing Interests] The authors declare that they have no competing financial
interests.
\item[Correspondence] Correspondence should be addressed to C. Titus Brown (ctb@msu.edu).
\end{addendum}

\pagebreak


\section*{Tables}

\begin{table}[h]
\caption{The total number of reads  in unfiltered, filtered (normalized
  and high abundance (HA) k-mer removal), and partitioned datasets and
  the computational resources required (memory and time).}
\resizebox{\columnwidth}{!}{%
\begin{tabular}{l c c c }

& Unfiltered & Filtered & Partitioned \\ 
& Reads (Mbp) & Reads (Mbp) & Reads (Mbp) \\
\hline
HMP Mock & 14,494,884 (1,136) & 8,656,520 (636) & 8,560,124 (631)  \\
HMP Mock Spike & 14,992,845 (1,137) & 8,189,928 (612) & 8,094,475 (607)  \\
HMP Mock Multispike & 17,010,607 (1,339) & 9,037,142  (702) & 8,930,840 (697)  \\
Iowa Corn & 1,810,630,781 (140,750)  & 1,406,361,241 (91,043) & 1,040,396,940 (77,603)  \\ 
Iowa Prairie & 3,303,375,485 (256,610) & 2,241,951,533 (144,962) & 1,696,187,797 (125,105)  \\ 
\\
 & Unfiltered (GB / h) & Filtered and Partitioned (GB / h) \\
HMP Mock &  4 / $<$2 & 4 / $<$2 \\
HMP Mock Spike & 4 / $<$2 & 4 /
$<$2 \\
HMP Mock Multispike &  4 / $<$2 &
4 / $<$2 \\
Iowa Corn & 188 / 83 &
234 / 120 \\ 
Iowa Prairie & 258 / 178 & 287 / 310 \\ \hline

\end{tabular}%
}
\label{data-summary}
\end{table}

\newpage

\begin{table}[h]
\caption{Assembly summary statistics (total contigs, total million bp
  assembly length, maximum contig size bp) of unfiltered (UF) and
  filtered (F) or filtered/partitioned (FP) datasets with Velvet (V)
  assembler.  Assembly for UF and FP datasets also shown for MetaIDBA
  (M) and SOAPdenovo(S) assemblers.  Iowa corn and prairie metagenomes
  could not be completed on unfiltered datasets.}
\resizebox{\columnwidth}{!}{%
\begin{tabular}{l c c c c}
& UF & F & FP & Assembler \\
\hline
HMP Mock & 29,063 / 38 / 146,795 & 30,082 / 35 / 90,497 & 30,115 / 35
/ 90,497 & V \\
HMP Mock & 24,300 / 36  / 86,445 & - & 27,475 / 36 / 96,041 & M \\
HMP Mock & 36,689 / 37 / 32,736 & - & 29,295 / 37 / 58,598 & S \\
Iowa corn & N/A & N/A & 1,862,962 / 912/ 20,234 & V \\
Iowa corn & N/A & N/A & 1,334,841 / 623 / 15,013 & M \\
Iowa corn & N/A & N/A & 1,542,436 / 675 / 15,075 & S \\
Iowa prairie & N/A & N/A & 3,120,263 / 1,510 / 9,397 & V \\
Iowa prairie & N/A & N/A & 2,102,163 / 998 / 7,206 & M \\
Iowa prairie & N/A & N/A & 2,599,767 / 1,145 / 5,423 & S \\
\end{tabular}%
}
\label{assembly-summary}
\end{table}

\newpage

\begin{table}[h]
\caption{Assembly comparisons of unfiltered (UF) and filtered (F) or
  filtered/partitioned (FP) HMP mock datasets using different
  assemblers (Velvet (V), MetaIDBA (M) and SOAPdenovo (S)).  Assembly
  content similarity is based on the fraction of alignment of
  assemblies and similarly, the coverage of reference genomes is based
  on the alignment of assembled contigs to reference genomes (RG).}
\begin{tabular}{l c c c}
Assembly Comparison & Percent Similarity & RG Coverage & Assembler \\
\hline
UF vs. F & 95\% & 43.3\% / 44.5\% & V \\
UF vs. FP & 95\% & 43.3\% / 44.4\% & V\\
UF vs. FP & 93\% & 46.5\% / 45.4\% & M\\ 
UF vs. FP & 98\% &  46.2\% / 46.4\% & S\\
\end{tabular}
\label{assembly-compare}
\end{table}

\newpage

\begin{table}[h]
\caption{Fraction of single-end (SE) and paired-end (PE) reads mapped
  to Iowa corn and prairie Velvet assemblies.}
\begin{tabular}{l c c}
 & Iowa Corn Assembly & Iowa Prairie Assemby \\
 \hline
Total Unfiltered Reads	& 1,810,630,781	& 3,303,375,485\\
Total Unfiltered SE Reads &	141,517,075 &	358,817,057\\
SE aligned 1 time	& 11,368,837	& 32,539,726\\
SE aligned $>$ 1 time	& 562,637	& 1,437,284\\
\% SE Aligned & 	8.43\% &	9.47\% \\
Total Unfiltered PE Reads & 	834,556,853	& 1,472,279,214\\
PE aligned 1 time	& 54,731,320	& 110,353,902\\
PE aligned $>$ 1 time	&1,993,902	 & 3,133,710\\
\% PE Aligned Disconcordantly	 & 0.47\% &	 0.63\%\\
\% PE Aligned	& 9.68\%	& 11.20\%\\
\end{tabular}
\label{read-map}
\end{table}

\newpage

\section*{Figures}

\begin{figure}[ht]
\center{\includegraphics[width=\textwidth,height=\textheight,keepaspectratio]
{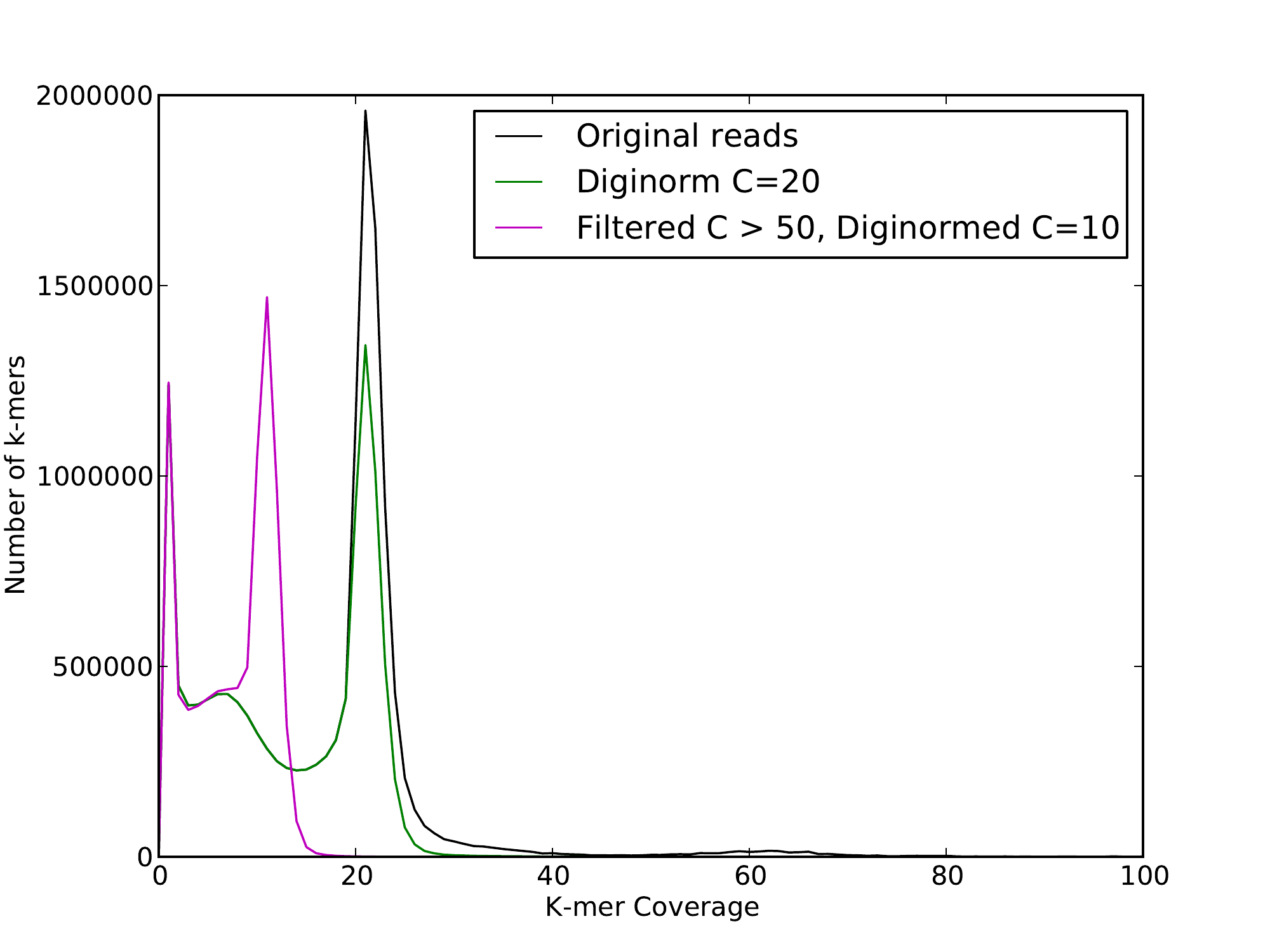}}
\caption{K-mer coverage of HMP mock community dataset before and
  after filtering approaches.}
\label{kmercoverage}
\end{figure}

\newpage

\begin{figure}[ht]
\center{\includegraphics[width=\textwidth,height=\textheight,keepaspectratio]
{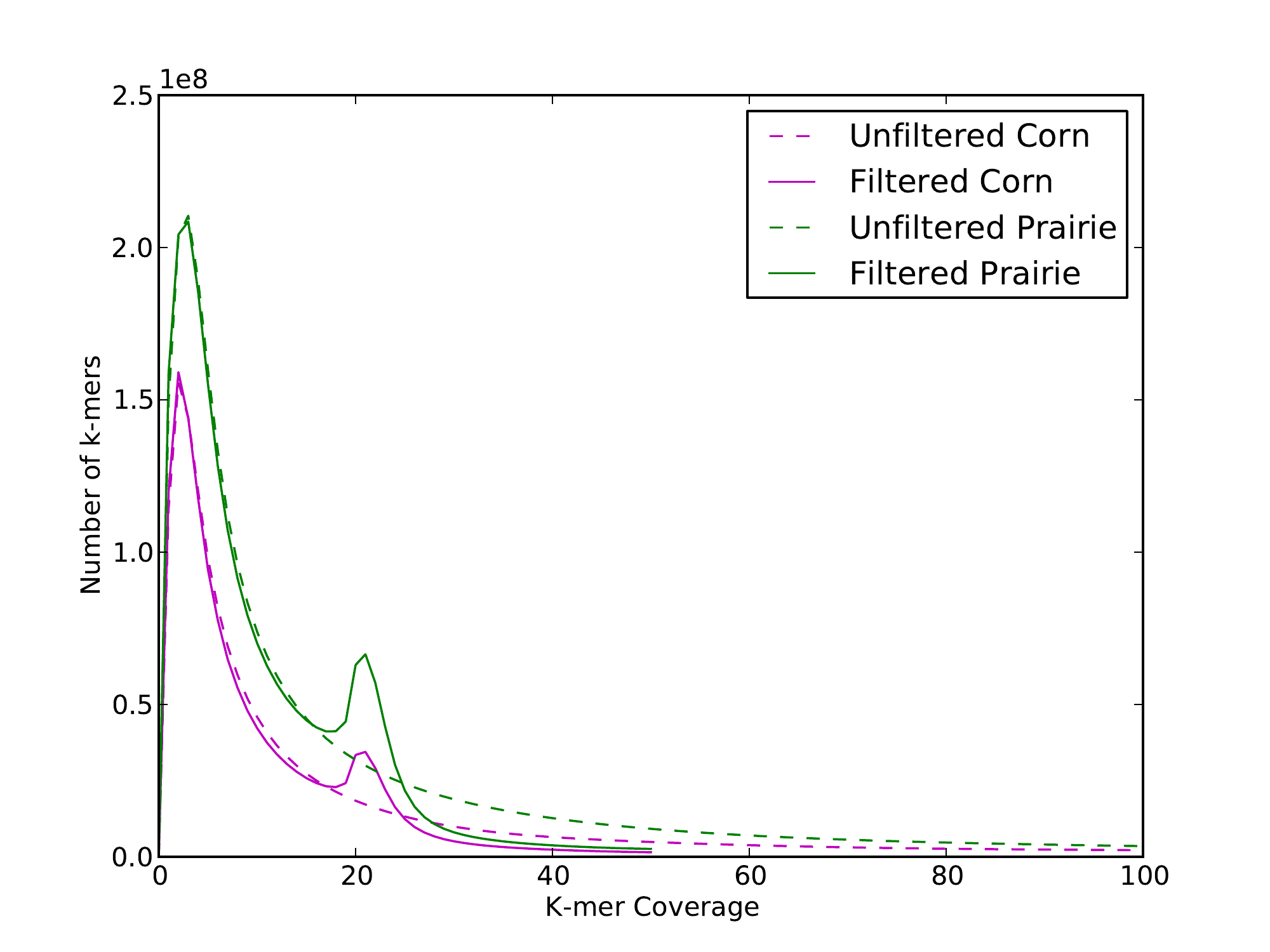}}
\caption{K-mer coverage of Iowa corn and prairie metagenomes before
  and after filtering approaches.}
\label{diginormcoverage}
\end{figure}

\newpage

\begin{figure}[ht]
\center{\includegraphics[width=\textwidth,height=\textheight,keepaspectratio]
  {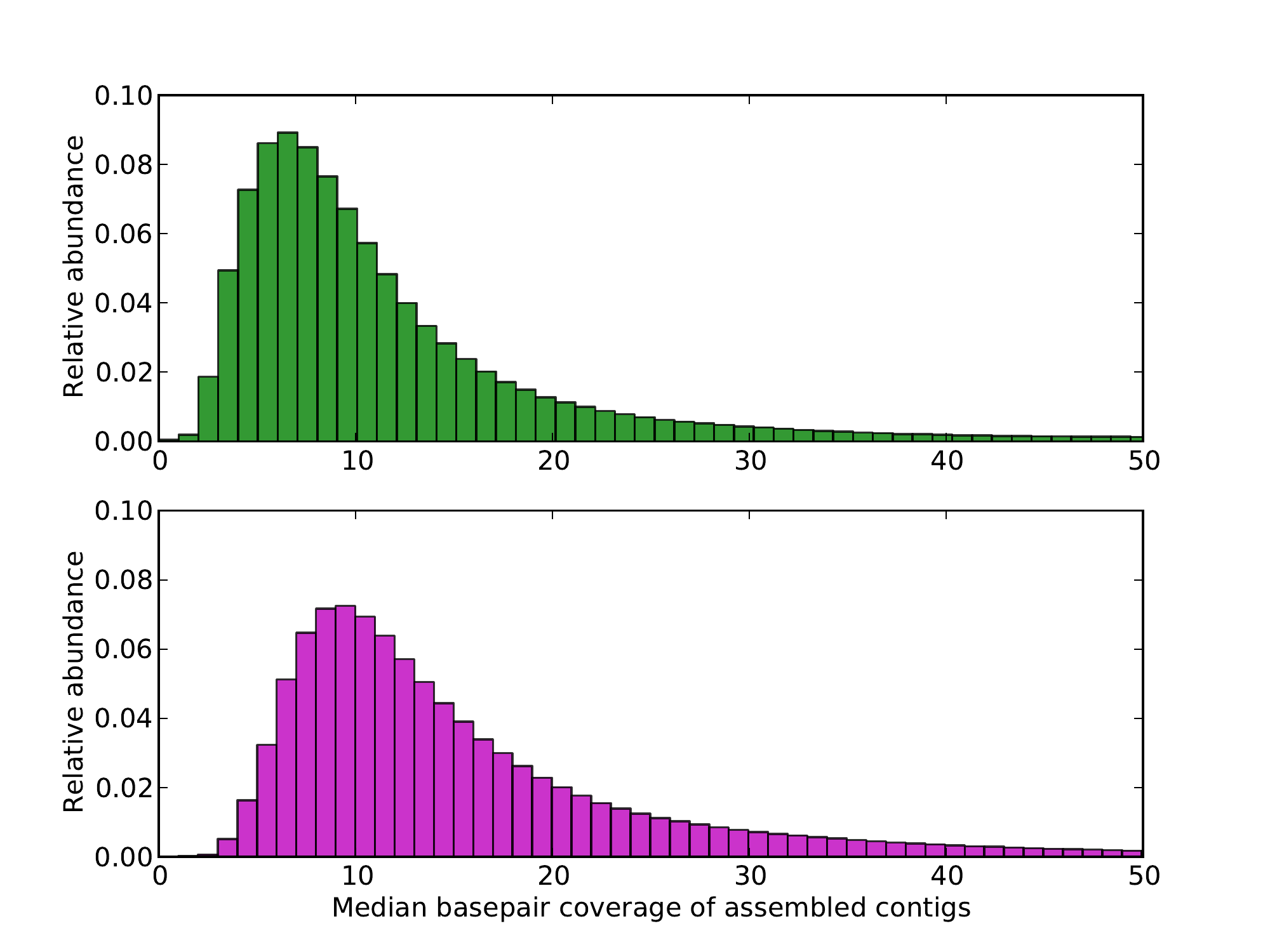}}
\caption{Coverage (median basepair recovered) distribution of assembled contigs
  from Iowa corn soil (top) and Iowa prairie soil (bottom) metagenomes.}
\label{soilassemblycoverage}
\end{figure}

\newpage

\begin{figure}[ht]
\center{\includegraphics[width=\textwidth,height=\textheight,keepaspectratio]
  {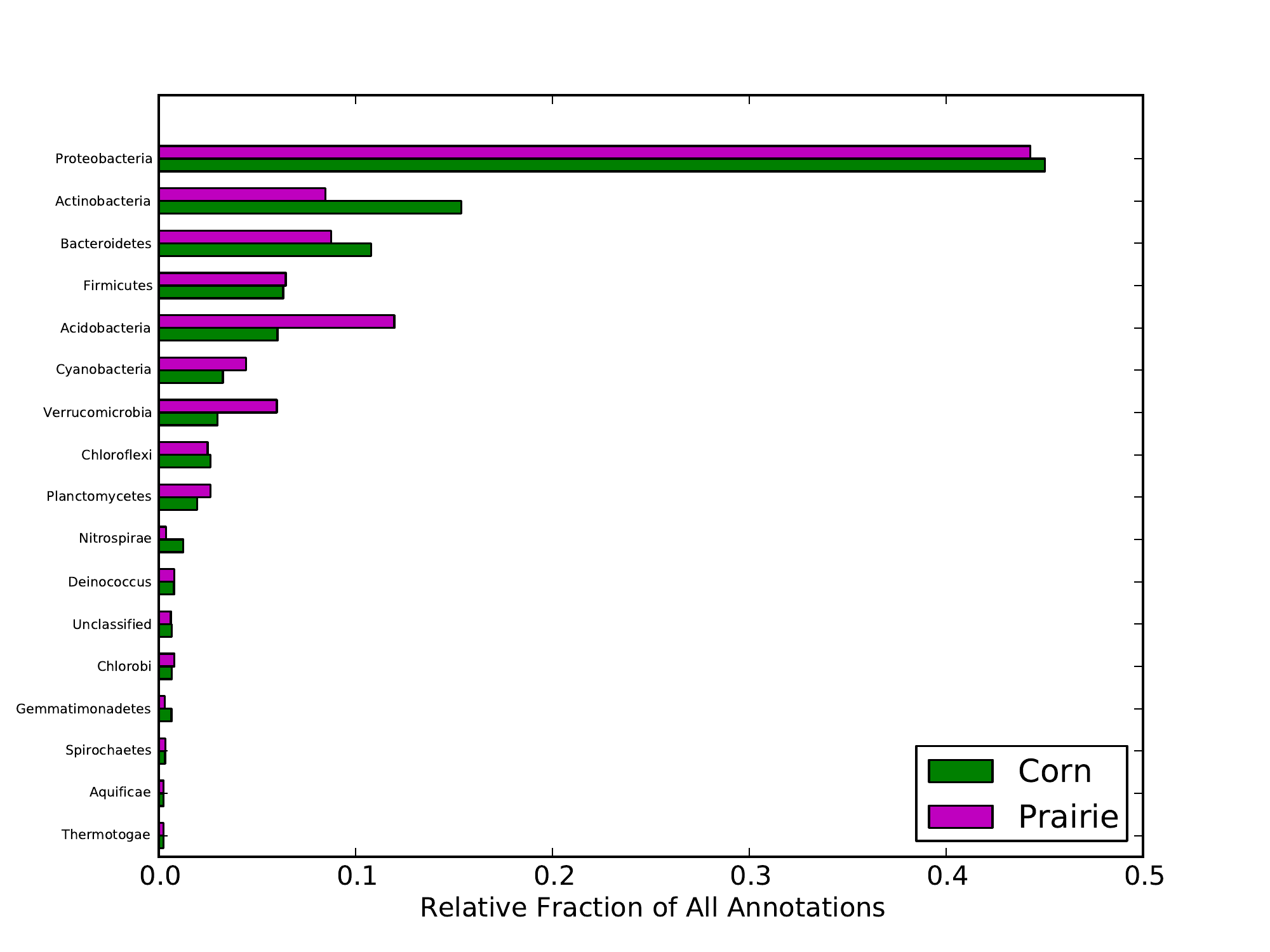}}
\caption{Phylogenetic distribution from SEED subsystem annotations for
  Iowa corn and prairie metagenomes.}
\label{phyla}
\end{figure}

\newpage

\begin{figure}[ht]
\center{\includegraphics[width=\textwidth,height=\textheight,keepaspectratio]
  {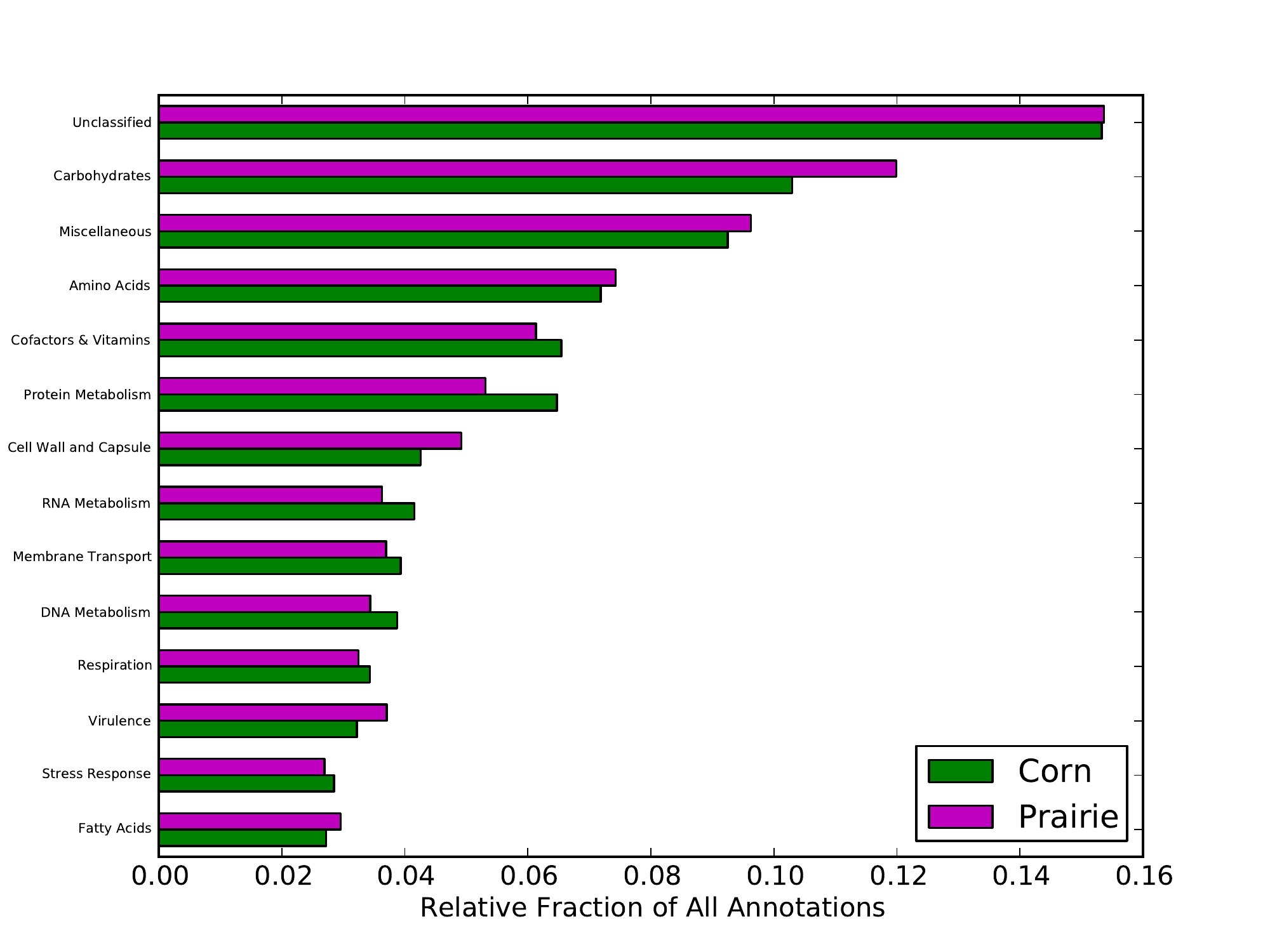}}
\caption{Functional distribution from SEED subsystem annotations for
  Iowa corn and prairie metagenomes.}
\label{subsystem}
\end{figure}

\newpage
\newpage

\begin{figure}[ht!]
\center{\includegraphics[scale=0.5]{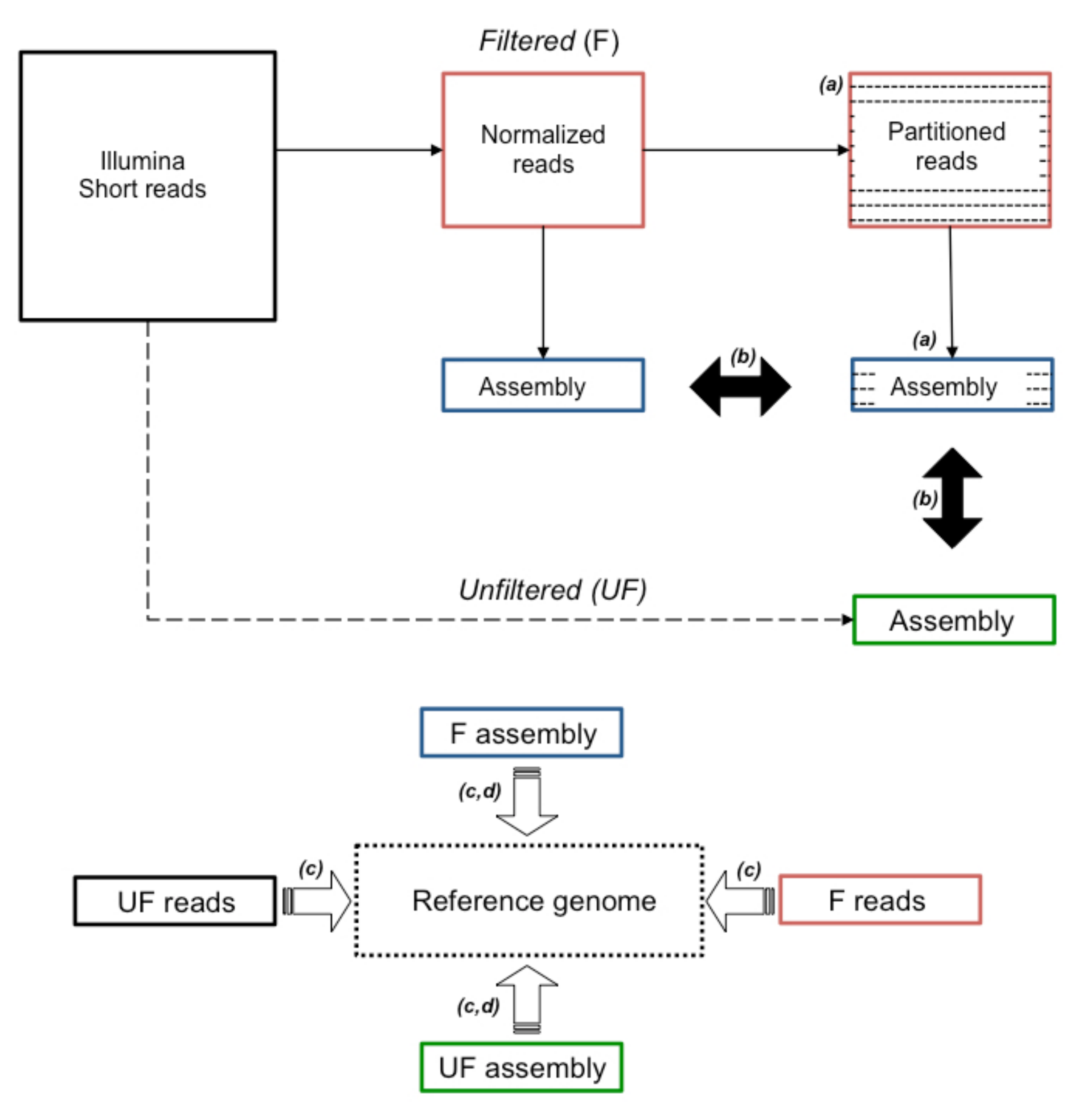}}
\caption{Flowchart describing methods for \emph{de novo} metagenomic assembly.   Using the HMP
mock community dataset, alternative approaches for data reduction and assembly were compared.  
(a) Disconnected subgraphs of the assembly graph were partitioned.  Most connected subgraphs contained 
reads and contigs aligning to distinct genomes (Supplementary Fig. 6). (b) The genomic content of all assemblies were 
found to be comparable in genomic content.  (c)  Reads and assembled contigs could be aligned to reference 
genomes to determine effectiveness of recovery. (d) The abundance of contigs (based on read mapping) could 
be compared to estimated abundances of corresponding reference genomes.} 
\label{flowchart}
\end{figure}

\newpage

\newpage
\clearpage

\section*{Online Methods}

\section*{Assembly Pipeline}

The entire assembly pipeline for the mock community is described in
detail in an IPython notebook available for download at 
\emph{http://nbviewer.ipython.org/urls/raw.github.com/ngs-docs/ngs-notebooks/
master/ngs-70-hmp-diginorm.ipynb}
and 
\emph{http://nbviewer.ipython.org/urls/raw.github.com/ngs-docs/ngs-notebooks/
master/ngs-71-hmp-diginorm.ipynb}.  Soil assembly was performed with
the same pipeline and parameter changes as described in Supplementary Information.
The annotated metagenome for Iowa corn can be found at
\emph{http://metagenomics.anl.gov/linkin.cgi?metagenome=4504797.3} and 
Iowa prairie at \emph{http://metagenomics.anl.gov/linkin.cgi?metagenome=4504798.3}.

\section*{Statistical Methods}
The reference-based abundance (from reads mapped to reference genomes)
and assembly-based abundance (from reads mapped to contigs) of genomes
were compared.  Using a one-directional, paired t-test of squared
deviations, the abundance estimates of the unfiltered and filtered
assemblies were compared.  The mean and standard deviation of 
the abundances of unfiltered contigs, filtered contigs, and reference genes
were 6.8 +/- 7.1, 8.1 +/- 7.7, and 7.8 +/- 5.2, respectively.  We expected the 
filtered assembly to have increased accuracy due to a reduction of errors (e.g. normalization
and high abundance filtering) and used a one-sided t-test which
indicated that abundance estimations from the filtered assembly were
significantly closer to predicted abundances from reference genomes
(n=28,652, p-value of 0.032).

\newpage

\bibliography{assembly-paper}

\end{document}


\maketitle
\section*{Summary of approaches used on mock community dataset}

The HMP mock community dataset and its available draft reference
genomes were used to evaluate our approaches towards data reduction
and partitioning for \emph{de novo} metagenomic assembly.  Reads of
the mock community dataset were initially digitally normalized to a
coverage threshold of 20 (as previously described in \cite{browndiginorm}),
reducing the total number of reads from 14 to 11 million.
Additionally, to remove possible sequencing artifacts associated with
high coverage sequences (previously described in Howe et al., in preparation),
highly-abundant sequences (20-mers present at coverage greater than
50-fold) were filtered and the dataset was further normalized to a
coverage of 10, resulting in a total of 9 million reads (Fig.
3).  Finally, the remaining reads were divided into
disconnected sets of reads resulting in a total of 85,818 partitions
containing greater than five reads (summarized in
Table 1).

\section*{Supplementary Methods}

\subsection*{Datasets}
In this study, we examined two large soil metagenomes generated from
soils collected from Iowa corn and native prairie soils.  Sequencing
was performed at the DOE Joint Genome Institute (Walnut Creek, CA).
Reads were quality trimmed at where Phred scores indicated a score of
'2'.  The total quality-trimmed reads in the Iowa corn and prairie
datasets were 1.8 million and 3.3 million, respectively.  We also
include a human gut mock community dataset (combined from SRA
SRX055381 and SRX055380).  For this mock community dataset, DNA from
bacterial isolates originally recovered from within or on the human
body was mixed together at staggered concentrations (over 5 orders of
magnitude based on genomic DNA concentrations) and sequenced.  The
mock community dataset originally contained 14.5 million reads.

To evaluate our approaches, we added simulated reads from either a
single E. Coli (str. K-12 substr DH10B) or five E. coli strains (K-12
substr DH10B, E24377, O147:H7 str. EC4115, UMN026, SE15) into select
metagenomes.  We computationally generated 100 bp reads from each
reference genome to a coverage of 10x and with a 2\% error rate and
subsequently randomly shuffled these reads with select datasets.

\subsubsection*{Estimation of assembly requirements for soil metagenomes}
Subsets of the Iowa corn metagenome were assembled with the Velvet
assembler (v1.2.07) with the following parameters: velveth K=45,
-short and velvetg -exp\_cov auto -cov\_cutoff auto, -scaffolding no.
The time and memory for each assembly was estimated up to a maximum of
150 hours and 100 GB.

\subsubsection*{Digital normalization}
Digital normalization was previously describe in \cite{browndiginorm}.  
For the mock
community dataset, digital normalization was performed with the
following parameters: K=20, coverage=20, and Bloom filter size = 1 GB
x 4.  For Iowa corn metagenome, digital normalization parameters were
as follows: K=20, coverage=20, and Bloom filter size = 48 GB x 4.
Similar parameters were used for the Iowa prairie metagenome, with the
exception that the Bloom filter size was 60 GB x 4.

\subsubsection*{Removal of high abundance sequences}
To eliminate known sequencing artifacts in Illumina metagenomes
(previously described in Howe et al., in preparation), high abundance 
sequences (coverage
greater than 50) were removed using the count-min-sketch datastructure
used for digital normalization.  For the relatively high coverage mock
community dataset, filtered reads were subsequently normalized to a
coverage of 10 (K=20, bloom filter size = 1 GB x 4).

\subsubsection*{Partitioning and \emph{de novo} assembly of disconnected reads}
Normalized and filtered datasets were loaded into a probabilistic
representation of the assembly graph as described in \cite{Pell:2012cq}, and
disconnected partitions of the resulting graph were separated.
Partitions containing less than five reads were discarded.  Each
partition was subsequently assembled using the Velvet assembler with
the same setting as described above, with the exception that K=35-59
and shortPaired setting was used for paired end reads.  The resulting
contigs greater than 300 bp from multiple-K assemblies were
dereplicated with CD-HIT (\cite{Fu:2012jk}, 99\% similarity) and merged with
Minimus2 \cite{Sommer:2007p1253}.

\subsection*{Comparing coverage of reference genomes by reads}
Reads in the HMP mock unfiltered and filtered datasets were mapped
back to originating genomes using default settings in Bowtie2
\cite{bowtie}.  For cases where reads could be mapped back to multiple
genomes, a single genome was randomly selected to be identified with
each read.  Sequencing coverage was estimated for the whole genome as
the median base pair coverage for all base pairs in the reference
genome.

\subsection*{Read coverage by assemblies}
All quality trimmed reads for Iowa corn and prairie were aligned with
assembled contigs (length greater than 300 bp) using default
parameters in Bowtie2 \cite{bowtie}.  Paired end reads were evaluated according to
concordance with paired end library preparation (i.e. paired end reads
on opposite DNA strands) and the alignment of both pairs of reads to
an assembled contig.  The base pair coverage of each contig was
estimated with the median base pair coverage of all reads across the
length of the contig.  Additionally, for each position in a contig
(with the exception of the external 100 bp on each end), the
percentage of the mapped consensus base pairs was calculated.  The
fraction of positions with greater than 95\% base consensus was
calculated to estimate the presence of polymorphisms within the
assembled contig.

\subsection*{Annotation of assemblies}
Assembled contigs and their corresponding median bp coverage for the
Iowa corn and prairie metagenomes were upload into MG-RAST annotation
pipeline \cite{Meyer:2008db} and are available on MG-RAST as 4504979.3 (Iowa corn)
and 4504798.3 (Iowa Prairie).  The resulting MG-RAST blat annotations
were compared to the M5NR database using a maximum e-value of 1e-5, a
minimum identity of 60\%, and a minimum alignment length of 15 aa .
Both the phylogenetic distribution of bacteria (phyla) and functional
distibution of subsystems were compared between the Iowa and corn
metagenomes.
  
\subsection*{Comparing assemblies}
Resulting assemblies (contigs greater than 300 bp) were compared using
the total number of contigs, assembly length, and maximum contig size
for each assembly.  Assemblies were also aligned to each other using
blastn and the resulting coverage of each assembly was calculated.  In
the case of the mock community, the resulting assemblies were also
aligned to sequenced draft genomes of the original isolates and, if
applicable, spiked reference genomes. Abundance of assembled contigs
and reference genomes were estimated by mapping raw reads with Bowtie
(allowing up to 2 mismatches for a match).  The median base pair
coverage was used to estimate abundances.  Associated assembled
contigs (greater than 300 bp) from the unfiltered and filtered
(digital normalized) assemblies were identified using a blastn
alignment (requiring E-value cutoff of 1e-5).  Contigs were associated
with reference genomes through an identical alignment approach.

The reference-based abundance (from reads mapped to reference genomes)
and assembly-based abundance (from reads mapped to contigs) of genomes
were compared.  Using a one-directional, paired t-test of squared
deviations, the abundance estimates of the unfiltered and filtered
assemblies were compared.  We expected the filtered assembly to have
increased accuracy due to a reduction of errors (e.g. normalization
and high abundance filtering) and used a one-sided t-test which
indicated that abundance estimations from the filtered assembly were
significantly closer to predicted abundances from reference genomes
(p-value of 0.032).

Annotations against the M5NR database were obtained through the
MG-RAST annotation pipeline.  The phylogenetic and functional
distribution of SEED subsystems between the Iowa corn and prairie
metagenome were compared (Fig. 6 and 7).  For each subsystem, the
relative abundance of each subsystem was calculated and the ratio of
the fraction present in the Iowa corn and prairie was determined
(e.g., the relative abundance of a subsystem which was equally
represented in both corn and prairie metagenomes would equal 1).  To
estimate similarity among all subsystems and phyla, the following was
calculated: ((1 - ratio)$^2$)$^{0.5}$ where a value closer to 0
indicates higher similarity.  Overall, for the phylogenetic and
functional distribution of SEED annotations, this value was 0.35 +/-
0.57 and 0.10 +/- 0.08.



\section*{Supplementary Tables}

\clearpage
\renewcommand{\figurename}{Supplementary Figure}
\renewcommand{\tablename}{Supplementary Table}
\begin{landscape}
\begin{table}
\caption{HMP mock dataset reference genomes estimated sequencing depth
  (median bp coverage of reads), number of partitions, total length
  (bp), coverage of reference genomes by unfiltered reads (UF Cov),
  coverage of reference genomes by filtered reads (F Cov), coverage of
  reference genomes by unfiltered assembled contigs (UFA Cov), and
  coverage of reference genomes by filtered assembled contigs (FA
  Cov).}
\begin{tabular}{l c c c c c c c}
\hline Reference Genome & Coverage & No. Partitions & Length (bp) & UF
Cov (bp) & F Cov (bp) & UFA Cov & FA Cov \\ \hline
gi\textbar{}32470588\textbar{}ref\textbar{}NC\_005008.1\textbar{} &
2,412 & 9 & 4,439 & 4,439 & 1,058 & 100 \% & 28 \% \\
gi\textbar{}32470581\textbar{}ref\textbar{}NC\_005007.1\textbar{} &
549 & 16 & 4,679 & 4,679 & 4,585 & 100 \% & 77 \% \\
gi\textbar{}32470520\textbar{}ref\textbar{}NC\_005003.1\textbar{} &
533 & 21 & 6,585 & 6,585 & 6,441 & 100 \% & 64 \% \\
gi\textbar{}32470572\textbar{}ref\textbar{}NC\_005006.1\textbar{} &
253 & 2 & 8,007 & 8,004 & 7,953 & 100 \% & 100 \% \\
gi\textbar{}32470532\textbar{}ref\textbar{}NC\_005004.1\textbar{} &
112 & 52 & 24,365 & 24,358 & 24,291 & 100 \% & 83 \% \\
gi\textbar{}126640109\textbar{}ref\textbar{}NC\_009084.1\textbar{} &
85 & 3 & 11,302 & 11,295 & 11,270 & 100 \% & 100 \% \\
gi\textbar{}32470555\textbar{}ref\textbar{}NC\_005005.1\textbar{} & 74
& 12 & 17,261 & 17,202 & 17,180 & 100 \% & 100 \% \\
gi\textbar{}10957398\textbar{}ref\textbar{}NC\_000958.1\textbar{} & 71
& 73 & 177,466 & 177,261 & 174,614 & 100 \% & 95 \% \\
gi\textbar{}10957530\textbar{}ref\textbar{}NC\_000959.1\textbar{} & 52
& 37 & 45,704 & 44,974 & 43,557 & 100 \% & 92 \% \\
gi\textbar{}126640097\textbar{}ref\textbar{}NC\_009083.1\textbar{} &
48 & 2 & 13,408 & 13,405 & 13,383 & 100 \% & 100 \% \\
gi\textbar{}15807672\textbar{}ref\textbar{}NC\_001264.1\textbar{} & 40
& 63 & 412,348 & 410,970 & 403,553 & 100 \% & 99 \% \\
gi\textbar{}15805042\textbar{}ref\textbar{}NC\_001263.1\textbar{} & 32
& 546 & 2,648,638 & 2,634,512 & 2,589,566 & 100 \% & 99 \% \\
gi\textbar{}27466918\textbar{}ref\textbar{}NC\_004461.1\textbar{} & 30
& 476 & 2,499,279 & 2,498,081 & 2,492,248 & 100 \% & 98 \% \\
gi\textbar{}125654693\textbar{}ref\textbar{}NC\_009008.1\textbar{} &
29 & 14 & 37,100 & 36,585 & 33,250 & 94 \% & 96 \% \\
gi\textbar{}161508266\textbar{}ref\textbar{}NC\_010079.1\textbar{} &
29 & 442 & 2,872,915 & 2,298,758 & 2,157,196 & 100 \% & 92 \% \\
gi\textbar{}77404776\textbar{}ref\textbar{}NC\_007490.1\textbar{} & 27
& 27 & 100,828 & 99,385 & 93,550 & 100 \% & 96 \% \\
gi\textbar{}125654605\textbar{}ref\textbar{}NC\_009007.1\textbar{} &
24 & 92 & 114,045 & 108,526 & 97,860 & 100 \% & 96 \% \\
gi\textbar{}77404693\textbar{}ref\textbar{}NC\_007489.1\textbar{} & 18
& 12 & 105,284 & 102,212 & 96,169 & 100 \% & 99 \% \\
gi\textbar{}24378532\textbar{}ref\textbar{}NC\_004350.1\textbar{} & 16
& 131 & 2,030,921 & 2,029,376 & 2,025,544 & 100 \% & 99 \% \\
gi\textbar{}77404592\textbar{}ref\textbar{}NC\_007488.1\textbar{} & 13
& 30 & 114,178 & 103,351 & 93,637 & 100 \% & 99 \% \\
gi\textbar{}77461965\textbar{}ref\textbar{}NC\_007493.1\textbar{} & 13
& 628 & 3,188,609 & 2,919,441 & 2,681,855 & 100 \% & 99 \% \\
gi\textbar{}77464988\textbar{}ref\textbar{}NC\_007494.1\textbar{} & 13
& 262 & 943,016 & 862,781 & 788,626 & 100 \% & 98 \% \\
gi\textbar{}126640115\textbar{}ref\textbar{}NC\_009085.1\textbar{} &
11 & 683 & 3,976,747 & 3,939,190 & 3,936,208 & 99 \% & 99 \% \\
gi\textbar{}148642060\textbar{}ref\textbar{}NC\_009515.1\textbar{} & 9
& 552 & 1,853,160 & 1,828,231 & 1,826,639 & 99 \% & 98 \% \\
gi\textbar{}150002608\textbar{}ref\textbar{}NC\_009614.1\textbar{} & 7
& 7,751 & 5,163,189 & 4,899,622 & 4,896,808 & 81 \% & 82 \% \\
gi\textbar{}15644634\textbar{}ref\textbar{}NC\_000915.1\textbar{} & 6
& 2,888 & 1,667,867 & 1,581,502 & 1,581,024 & 78 \% & 79 \% \\
gi\textbar{}194172857\textbar{}ref\textbar{}NC\_003028.3\textbar{} & 6
& 4,123 & 2,160,842 & 2,047,832 & 2,037,347 & 78 \% & 78 \% \\
gi\textbar{}49175990\textbar{}ref\textbar{}NC\_000913.2\textbar{} & 6
& 5,913 & 4,639,675 & 4,080,605 & 4,074,119 & 84 \% & 85 \% \\
gi\textbar{}50841496\textbar{}ref\textbar{}NC\_006085.1\textbar{} & 6
& 6,459 & 2,560,265 & 2,169,547 & 2,169,056 & 59 \% & 64 \% \\
gi\textbar{}77358697\textbar{}ref\textbar{}NC\_003112.2\textbar{} & 4
& 9,269 & 2,272,360 & 1,655,023 & 1,626,301 & 28 \% & 33 \% \\
\end{tabular}
\label{ref-summary}
\end{table}
\end{landscape}


\clearpage
\section*{Supplementary Figures}

\begin{figure}[ht]
\center{\includegraphics[width=\textwidth,height=\textheight,keepaspectratio]
{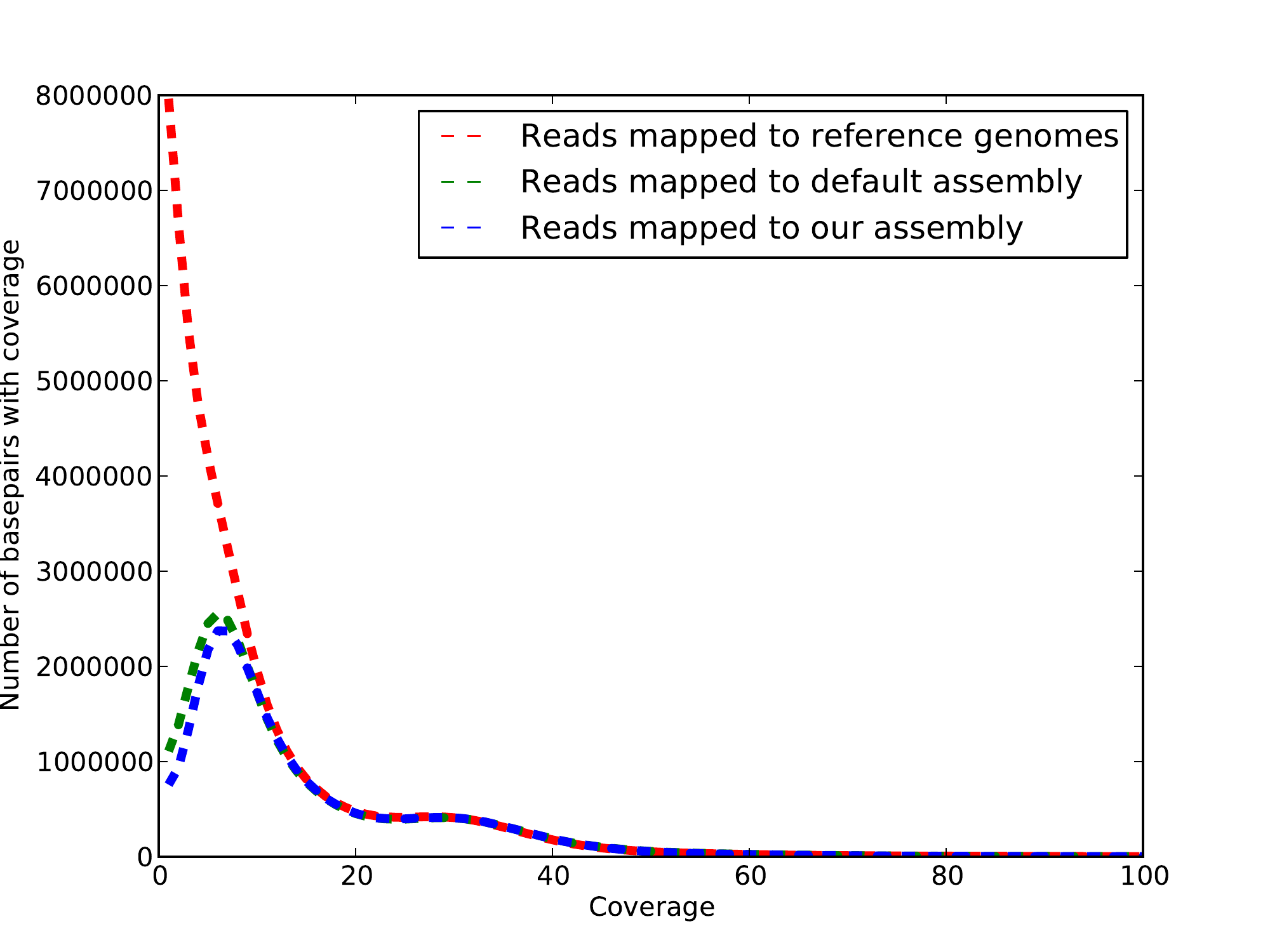}}
\caption{Number of basepairs with specified coverage for reads which
  map to reference genomes and unfiltered and filtered assembled
  contigs greater than 300 bp.}
\label{coveragehmp}
\end{figure}

\begin{figure}[ht]
\center{\includegraphics[scale=0.8]{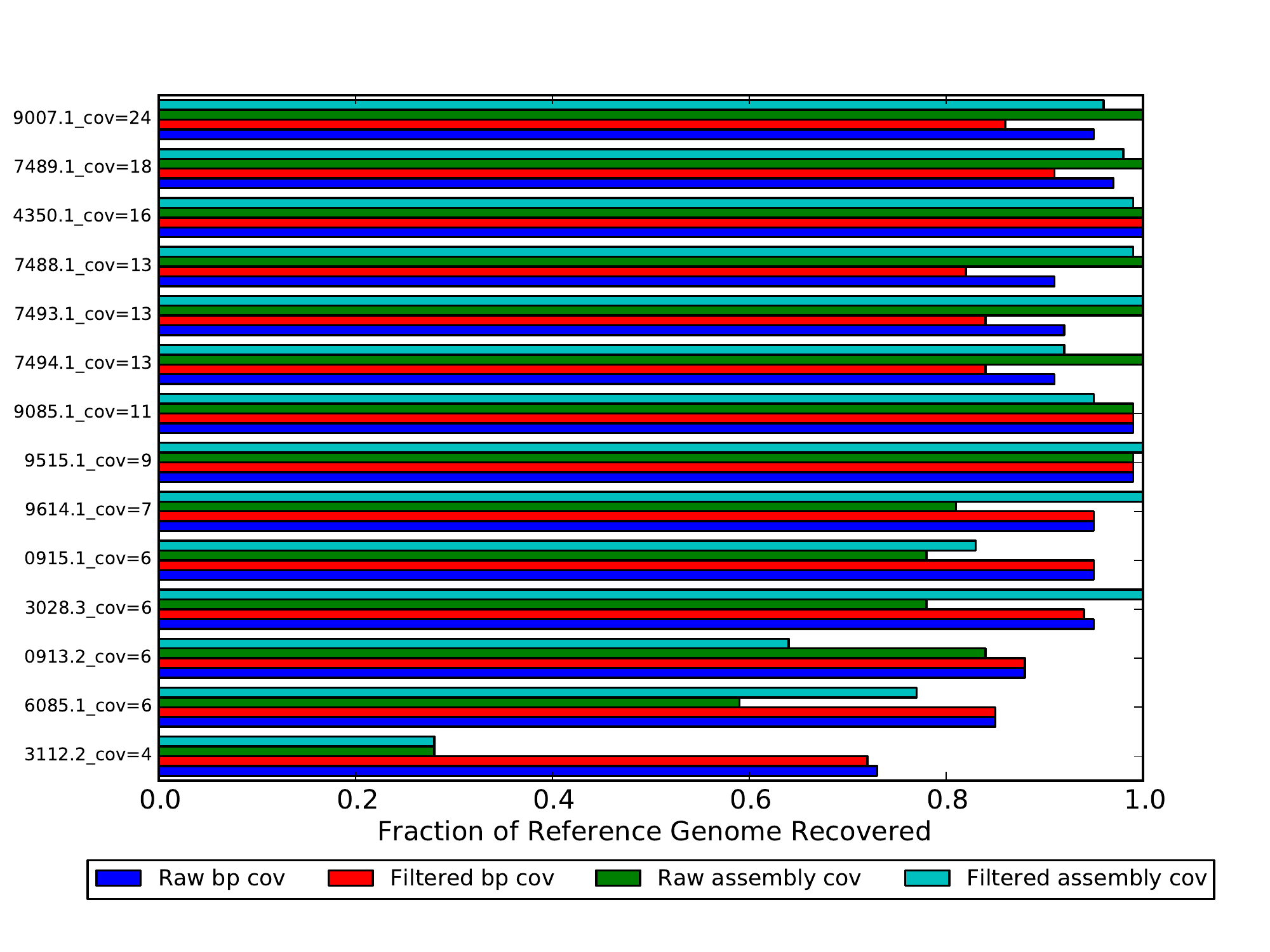}}
\caption{Coverage of reference genomes by unfiltered and filtered
  assembled contigs and unfiltered and filtered reads.}
\label{coverage1}
\end{figure}

\begin{figure}[ht]
\center{\includegraphics[scale=0.8]{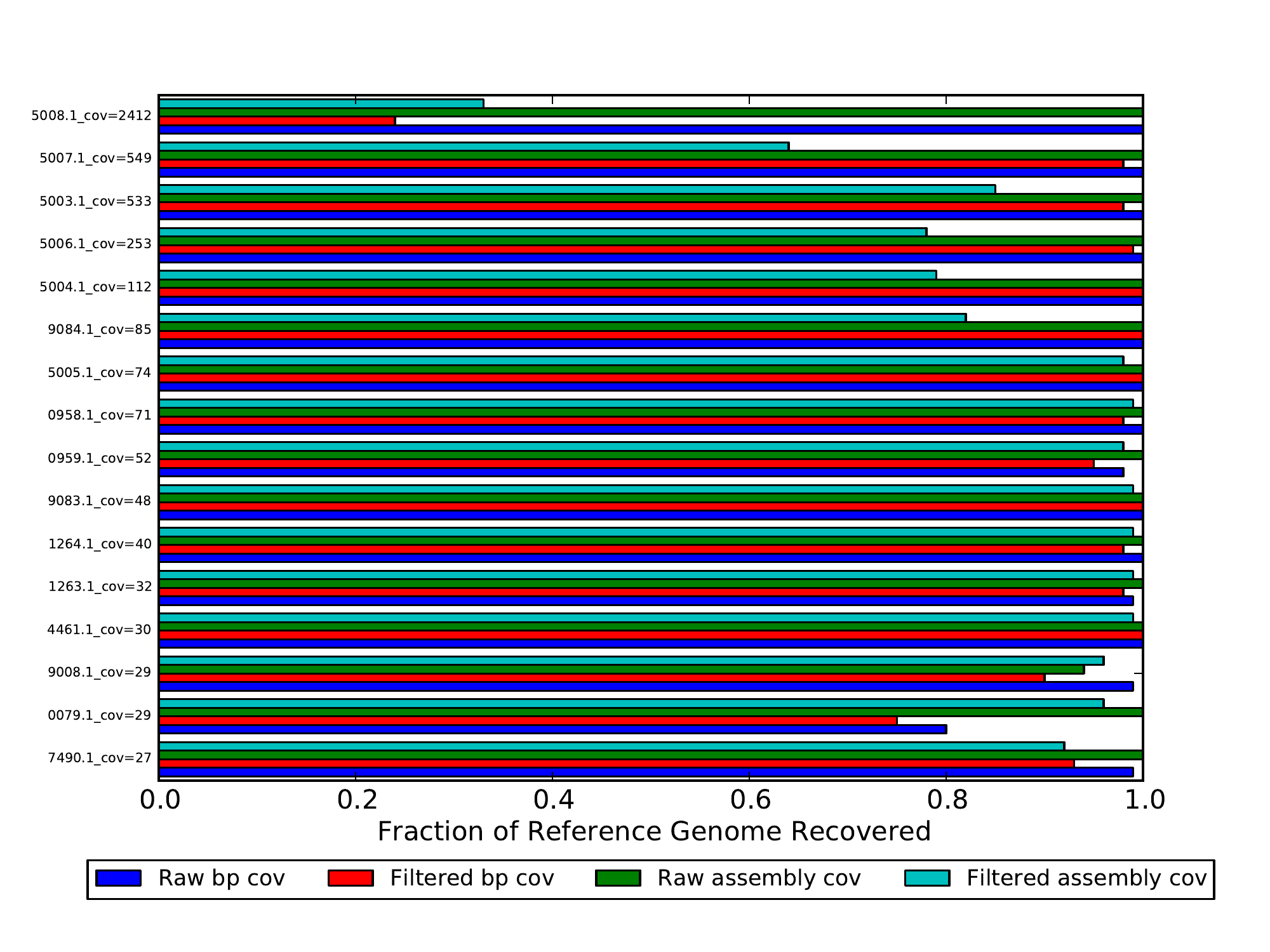}}
\caption{Coverage of reference genomes by unfiltered and filtered
  assembled contigs and unfiltered and filtered reads.}
\label{coverage2}
\end{figure}

\begin{figure}[ht]
\center{\includegraphics[scale=0.5]{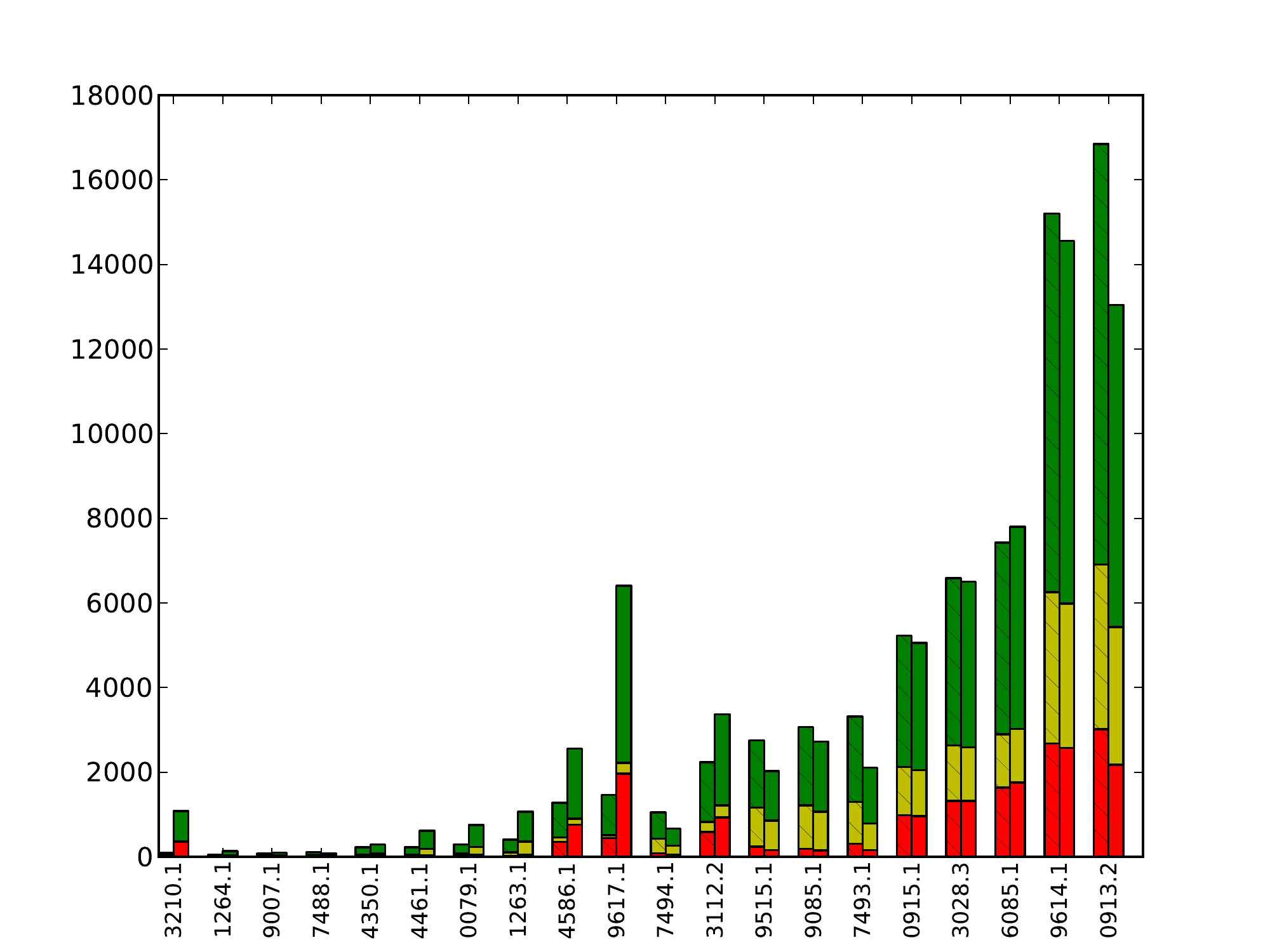}}
\caption{Total number of contigs for unfiltered (bars with hashed
  lines) and filtered (solid bars) for top twenty references with most
  assembled contigs (ranked by unfiltered assembly).  Red indicates
  contig lengths less than 500 bp, yellow indicates contig lengths
  between 500 bp and 3000 bp, and green indicates contig lengths
  greater than 5000 bp.  Reference genome IDs shown here are last 5
  digits of RefSeq ID.}
\label{contig-lengths}
\end{figure}

\begin{figure}[ht]
\center{\includegraphics[width=\textwidth,height=\textheight,keepaspectratio]{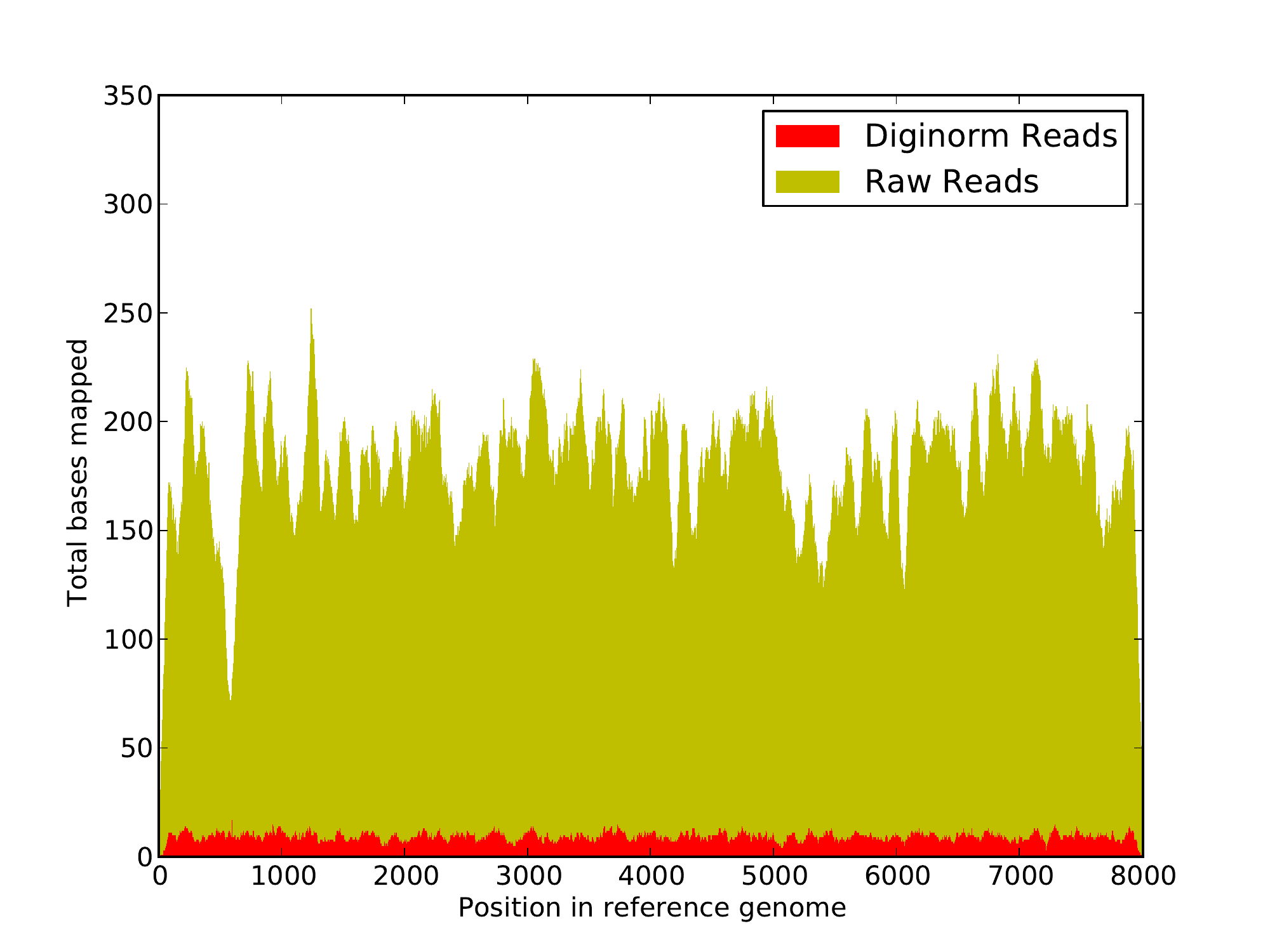}}
\caption{Alignment of reads (colored by originating partition) to
  reference genome NC\_00745901}
\label{diginormreference}
\end{figure}

\begin{figure}[ht]
\center{\includegraphics[width=\textwidth,height=\textheight,keepaspectratio]{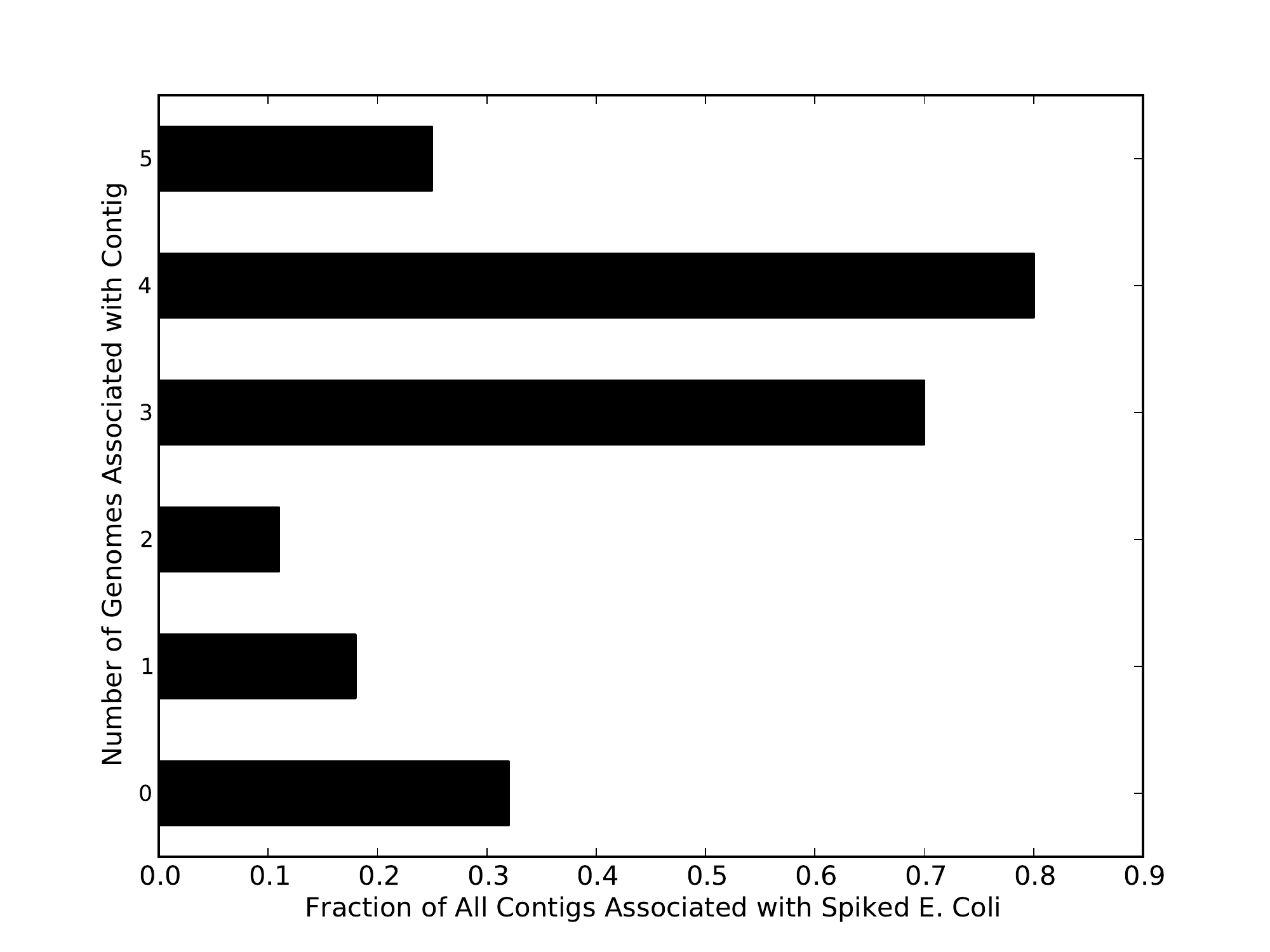}}
\caption{The fraction of assembled contigs assembled from partitions
  containing spiked \emph{E. coli} reads associated with 0 to five of
  the \emph{E. coli} reference genomes.  The large majority of contigs
  contain reads associated with multiple genomes or to no genome.}
\label{fractionassembled}
\end{figure}

\begin{figure}[ht]
\center{\includegraphics[width=\textwidth,height=\textheight,keepaspectratio]{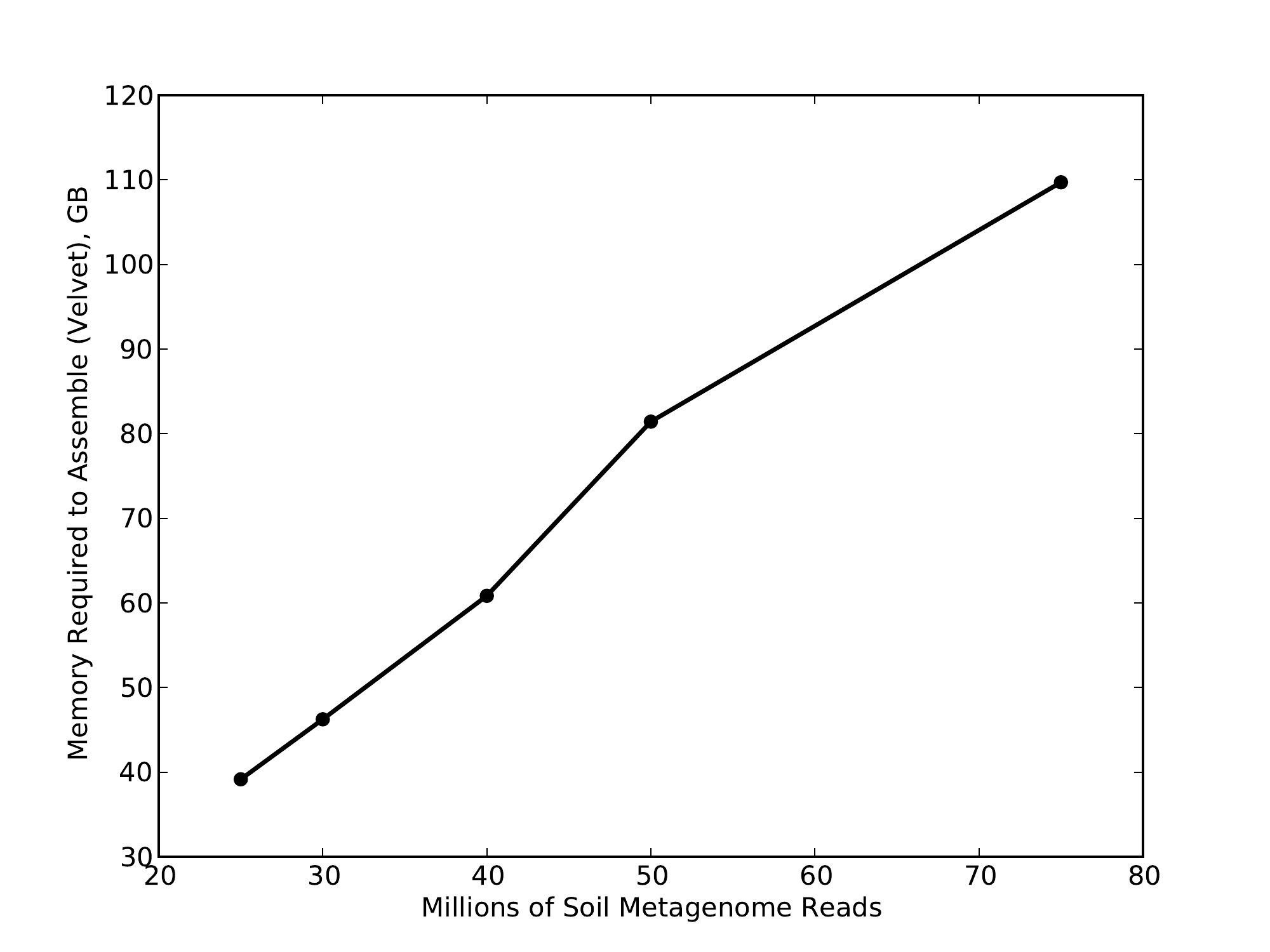}}
\caption{Memory requirements to assemble subsets of Iowa corn soil
  metagenome}
\label{memory}
\end{figure}

\begin{figure}[ht]
\center{\includegraphics[width=\textwidth,height=\textheight,keepaspectratio]{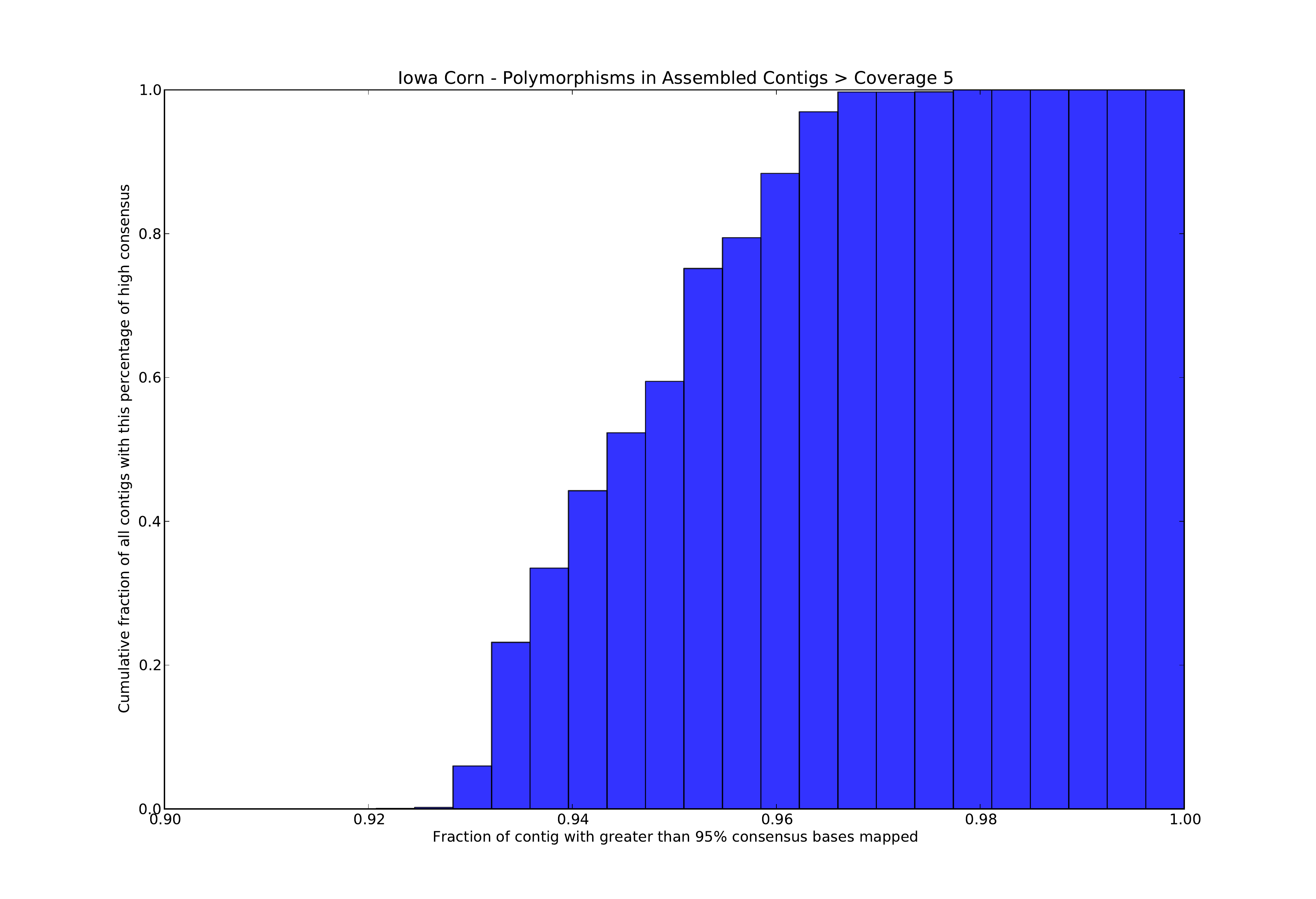}}
\caption{The presence of polymorphic sequences in assembled contigs of
  Iowa corn metagenome.}
\label{corn-poly}
\end{figure}

\begin{figure}[ht]
\center{\includegraphics[width=\textwidth,height=\textheight,keepaspectratio]{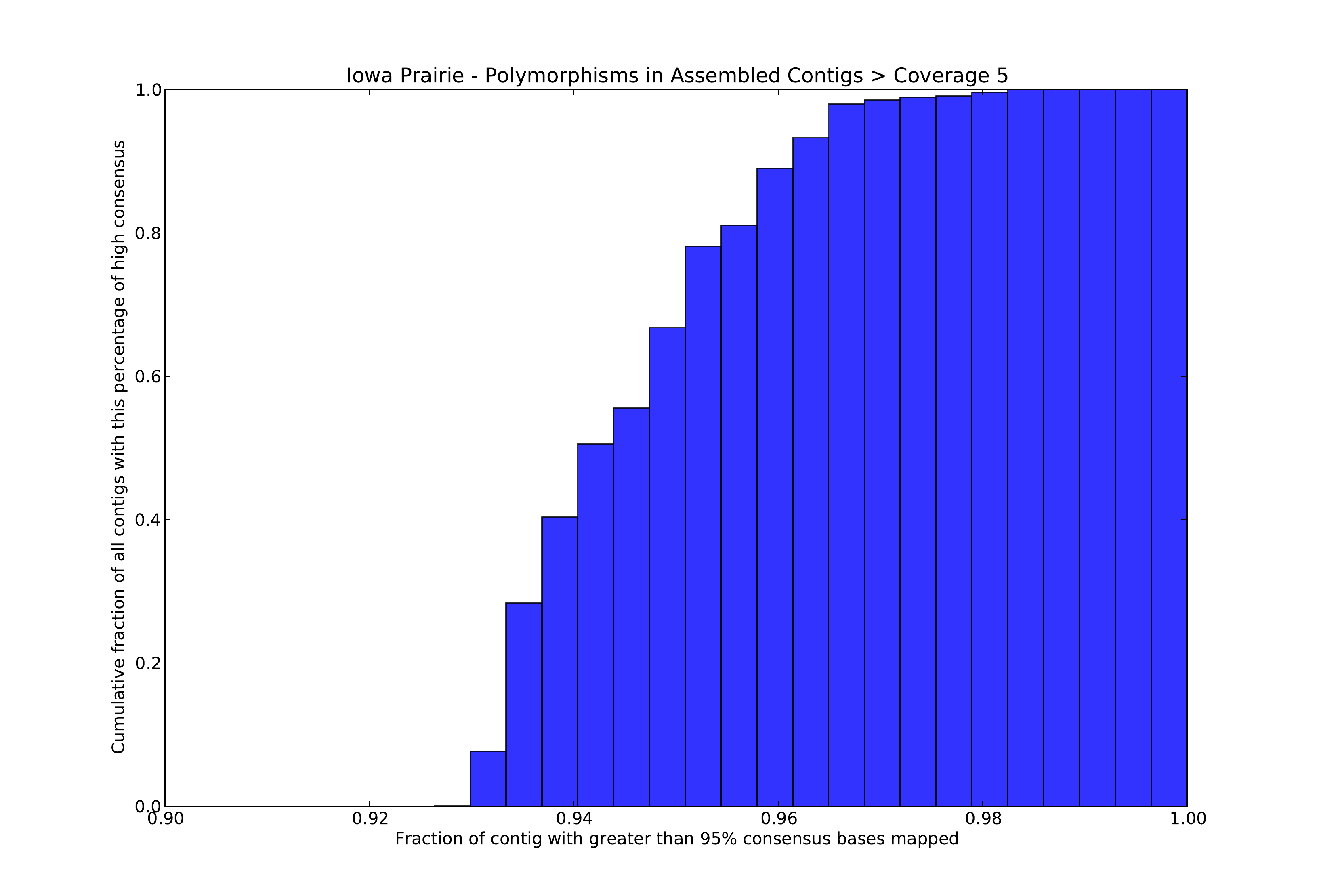}}
\caption{The presence of polymorphic sequences in assembled contigs of
  Iowa prairie metagenome.}
\label{prairie-poly}
\end{figure}

\clearpage

\section*{Supplementary References}
\bibliography{assembly-paper}